\documentclass[aps,showpacs,preprintnumbers,floats,amsmath,twocolumn]{revtex4}

\usepackage{graphicx}
\usepackage{dcolumn}
\usepackage{bm}
\usepackage{pdfpages}
\makeatother
\begin{document}

\title{Strange matter and strange stars in a thermodynamically self-consistent perturbation model with running coupling and running strange quark mass}
\author{J.~F.~Xu$^1$,
        G.~X.~Peng,$^{1,2}$\footnote{Correspondence to: gxpeng@ucas.ac.cn}
        F. Liu,$^3$
        D.~F. Hou,$^3$
        L.~W. Chen$^4$
        }
\affiliation{%
$^1$\mbox{School of Physics,
    University of Chinese Academy of Sciences, 380 Huaibeizhen, Beijing 101408, China}\\
$^2$\mbox{Theoretical Physics Center for Science Facilities,
    Institute of High Energy Physics,  Beijing 100049, China}\\
$^3$\mbox{Key Lab.\ of QLP, Institute\ of Particle Physics,
     Huazhong Normal University, Wuhan 430079, China}\\
$^4$Department of Physics and Astronomy, Shanghai Jiao Tong University, Shanghai 200240, China%
            }

\begin{abstract}
A quark model with running coupling and running strange quark mass,
which is thermodynamically self-consistent
at both high and lower densities, is presented and applied to study properties of
strange quark matter and structure of compact stars. An additional term
to the thermodynamic potential density is determined by
meeting the fundamental differential equation of thermodynamics.
It plays an important role in comparatively lower density and ignorable
at extremely high density, acting as a chemical-potential
dependent bag constant. In this thermodynamically enhanced perturtative QCD model,
strange quark matter still has the possibility of being absolutely stable, while
the pure quark star has a sharp surface
with a maximum mass as large as about 2 times the solar mass
and a maximum radius of about 11 kilometers.
\end{abstract}


\maketitle

\section{Introduction}
\label{intro}
In recent decades, strange quark matter (SQM) has been one of the
most interesting and significant topics in nuclear physics \cite{SQM2013}.
Early in 1970's, the possible existence of a deconfined phase
was proposed and studied \cite{Bodmer1971,Itoh1970,Chin1979,Terazawa1979}.
Specially in 1984, it was speculated, based on elementary symmetry considerations,
that SQM might be absolutely stable and thus have important consequences \cite{Witten1984}.
Soon after this it was shown in the MIT bag model that SQM is absolutely stable
for a reasonable range of QCD-related parameters \cite{Farhi1984}.
Since then a lot of works had been done on the properties and applications
\cite{Hou2015,Li2015,Isayev2015,Zhang2015,Chu2014a,Chu2014b%
,Paulucci2014,Chen2014,Sinha2013,Chang2013,Chen2012,Tang2011,Li2010,BaoT2008,WangQun2009}.

It is widely believed that the quantum chromodynamics (QCD) is the fundamental theory
of strong interactions, and in principle, one could do detailed
and comprehensive study on SQM by solving the motion
equations of quarks and gluons. Unfortunately, however,
QCD is, in fact, intractable in the nonperturbative regime presently. In particular
at finite baryon chemical potential, there is a notorious sign problem where the
Lattice Monte Carlo simulation is inaccessible \cite{Lat-sign}. Therefore,
effective phenomenological models play crucial roles to extract and figure out
the properties of strongly interacting matter.
In the past years, a number of models have been applied with interesting results,
such as the Nambu-Jona-Lasinio model \cite{NJL1961,XiaT2013},
the global color symmetry model \cite{ChangL2005},
the quasiparticle model \cite{Goloviznin1993,Peshier1994,Bluhm2005,Gardim2009,Banuur2007,Wen2013prd,Zong2013},
the mass-density-dependent model \cite{Fowler1981,Chakrabaty1989,Benvenuto1995,Peng2000a, Peng2000b,Wen2005},
the equivparticle model \cite{Peng2008,Xia2014},
the quark-cluster model \cite{Xu2003-2010}, and so on.

Thermodynamic consistency is a fundamental requirement
of phenomenological models \cite{Peng2000b}.
In many important cases, an additional term to the thermodynamic potential density is necessary to
maintain thermodynamic consistency. In an important version of the quasiparticle model,
for example, the additional term is needed in both the zero \cite{Schertler1997npa}
and finite temperature \cite{Gorenstein1995} cases.
In the equivparticle model with confinement by the density dependence of quark masses,
an additional term also appears in the thermodynamic potential density
to have full thermodynamic consistency \cite{Peng2008,Xia2014}.

Because of  asymptotic freedom, the perturbative calculation of QCD
is reliable at very high densities.
The thermodynamic potential density of cold quark matter was
calculated for massless quarks in Refs.~\cite{Freedman1977,Baluni1978,Toimela1985}.
These results were applied to study quark stars to the first
order in QCD coupling in Refs.~\cite{Alcock1986,Haensel1986},
to the second order in \cite{Fraga2001},
with finite-mass effect of strange quarks considered in Refs.\
\cite{Fraga2005,Fraga2006npa,Kurkela2010}.

The validity of a perturbative theory requires a small coupling with
which the perturbative series is obtained.
Different from quantum electrodynamics (QED), however, the
QCD coupling is running, i.e., it is not that small when the density is not
extremely high. In this case, one will meet thermodynamic problem
when one naively extends the applicable range of density \cite{Peng2005}.

One way to solve this problem is to add an additional term
to the thermodynamic potential density, similar to that of the
popular quasiparticle model \cite{Schertler1997jpg,
Wen2009,Wen2010,Wen2012} and the equivparticle model \cite{Peng2008,Xia2014}.
This way of extending the applicable range of the perburbative calculation
was shown to be reasonable for the one-flavor case \cite{Peng2005}.
It has also been shown, for the massless two-flavor case,
that the renormalization subtraction point should be taken
as a function of the summation of the biquadratic chemical potentials
while the additional term not only keeps the thermodynamics
self-consistent, but also produces reasonable results \cite{XuJF2014}.
In the present paper, we extend this thermodynamically
enhanced perturbative QCD (EPQ) model  to the actual SQM with massless
up (u) and down (d) quarks plus massive strange (s) quarks.
It is found that the additional term takes an important role
at lower densities, acting as a chemical-potential dependent bag constant.
The equation of state (EOS) of SQM becomes stiffer, and accordingly
the maximum mass of strange stars is as large as two times the solar mass.

The paper is organized as follows.
In the next section \ref{SecConvenTreat},
we give the conventional perturbative treatment and demonstrate
the thermodynamic inconsistency in its naive extension to lower densities.
Then in Sec.\ \ref{newTreat}, we determine
a chemical-potential dependent bag-like coupling constant which makes the thermodynamic
treatment self-consistent. After that
we study, respectively, the properties of SQM and the
structure of compact stars with the new
EPQ model in Sec. \ref{EOS-SQM} and \ref{struc-stars}.
Finally the section \ref{summary} is a short summary.

\section{The conventional perturbation model and inconsistency of its naive extrapolation}
\label{SecConvenTreat}

Let's start our work from the perturbative expansion
of the thermodynamic potential density of cold quark matter
with two-flavor massless light quarks plus one massive strange quark. According
to Eqs. (1) and (2) in Refs. \cite{Fraga2005,Fraga2006npa},
we have the perturbative contribution to the thermodynamic potential density as
\begin{equation}\label{Omegaquark}
\Omega^{\mathrm{pt}}=\Omega_\mathrm{u}+\Omega_\mathrm{d}+\Omega_\mathrm{s},
\end{equation}
where, to the first order, the contributions from massless up and down quarks are respectively
\begin{eqnarray}\label{Omegaud}
\Omega_\mathrm{u}
&=& -\frac{\mu_\mathrm{u}^4}{4\pi^2}(1-2\alpha) \ \
\mathrm{and}\
\Omega_\mathrm{d}
=
-\frac{\mu_\mathrm{d}^4}{4\pi^2}(1-2\alpha),
\end{eqnarray}
and that from the massive strange quarks is \cite{Fraga2005,Kurkela2010}
\begin{eqnarray}\label{Omegas}
\Omega_\mathrm{s}
&=&
 \frac{-1}{4\pi^2}
 \left[
   \mu_\mathrm{s}\nu_\mathrm{s}(\mu_\mathrm{s}^2-\frac{5}{2}m_\mathrm{s}^2) 
   +\frac{3}{2}m_\mathrm{s}^4\,\mbox{ach}\frac{\mu_\mathrm{s}}{m_\mathrm{s}}
 \right]
\nonumber\\
&&
+\frac{\alpha}{2\pi^2}
 \left[
  3\left(\mu_\mathrm{s}\nu_\mathrm{s}-m_\mathrm{s}^2\,\mbox{ach}
  \frac{\mu_\mathrm{s}}{m_\mathrm{s}}\right)^2 -2\nu_\mathrm{s}^4
 \right.
\nonumber\\
&& 
 \left.
 +m_\mathrm{s}^2\Big(6\ln\frac{u}{m_\mathrm{s}}+4\Big)
  \left(\mu_\mathrm{s}\nu_\mathrm{s}-m_\mathrm{s}^2\mbox{ach}
  \frac{\mu_\mathrm{s}}{m_\mathrm{s}}\right)
\right].
\end{eqnarray}
Here $\mu_\mathrm{u}$, $\mu_\mathrm{d}$, and $\mu_\mathrm{s}$ are the chemical potentials of up,
down, and strange quarks respectively, $u$ is the renormalization
subtraction point, $\alpha\equiv \alpha_s/\pi=g^2/(4\pi^2)$
is the running coupling, and $\mbox{ach}\,x\equiv \ln(x+\sqrt{x^2-1})$ is the
inverse hyperbolic cosine function. Because the mass of u and d
quarks is much smaller than that of s quarks,
we consider only the mass effect of strange quarks.
For simplicity, we have used the notation $\nu_\mathrm{s}\equiv\sqrt{\mu_\mathrm{s}^2-m_\mathrm{s}^2}$,
which can be regarded as the fermion momentum of s quarks.
Because the electron does not participate in the strong interactions,
its contribution to the thermodynamic potentail is then
\begin{equation}\label{Omegae}
\Omega_\mathrm{e}=-\frac{\mu_\mathrm{e}^4}{12\pi^2}.
\end{equation}

The number densities of u and d quarks and electrons are, respectively
\begin{equation}\label{nu-nd}
n_\mathrm{u}=\frac{\mu_\mathrm{u}^3}{\pi^2}(1-2\alpha), \ n_\mathrm{d}=\frac{\mu_\mathrm{d}^3}{\pi^2}(1-2\alpha), \
n_\mathrm{e}=\frac{\mu_\mathrm{e}^3}{3\pi^2},
\end{equation}
while that of s quarks is
\begin{equation}\label{ns}
n_\mathrm{s}
=
 \frac{\nu_\mathrm{s}^3}{\pi^2}
-\frac{2\alpha}{\pi^2}\nu_\mathrm{s}
 \left(
   \mu_\mathrm{s}\nu_\mathrm{s}+2m_\mathrm{s}^2-3m_\mathrm{s}^2\,\ln\frac{\mu_\mathrm{s}+\nu_\mathrm{s}}{u}
 \right).
\end{equation}

Here we should keep in mind that all terms with order in the coupling higher than unity
have been discarded because we assume the perburbative expression is merely valid to leading order.

The thermodynamic potential density of the whole system composed
of u, d, s quarks and electrons, given
by the sum of Eq. (\ref{Omegaud}) to Eq. (\ref{Omegae}), depends on the quark
chemical potentials $\mu_\mathrm{u}$, $\mu_\mathrm{d}$, $\mu_\mathrm{s}$, $\mu_\mathrm{e}$, and $u$\ both
explicitly and implicitly through the scale dependence of
$\alpha(u)$ and the mass $m_\mathrm{s}(u)$.

The running coupling $\alpha(u)$ and the running mass $m_\mathrm{s}(u)$ of strange quarks
are determined by the following renormalization group (RG)
equations
\begin{equation} \label{alphadif}
\frac{\mathrm{d}\alpha}{\mathrm{d}\ln u^2}
=-\sum_{j=0}^{{\cal N}-1}\beta_i\alpha^{i+2}\equiv \beta(\alpha),
\end{equation}
\vspace{-5mm}
\begin{equation} \label{msdif}
\frac{\mathrm{d}m_s}{\mathrm{d}\ln u^2}
=-\sum_{i=0}^{{\cal N}-1}\gamma_i\alpha^{i+1}
\equiv \gamma(\alpha),
\end{equation}
where ${\cal N}$ is the loop number, while the beta and gamma functions, $\beta(\alpha)$ and $\gamma(\alpha)$, are presently known to four-loop level,
given by the corresponding beta and gamma coefficients, i.e., $\beta_i$ and $\gamma_i$.
The original ones were given in the minimum subtraction scheme
or its modified version ($\overline{\mbox{MS}}$). The coupling
and masses given with these $\beta_i$ and $\gamma_i$ are,
in principle, not continuous at heavy quark thresholds.
In order to give a continuous coupling and a continuous
strange quark mass, the beta and gamma coefficients
should be recombined. For the number of colors $N_\mathrm{c}=3$,
these coefficients are provided in the appendix A.
Comparing these coefficients with the original beta
and gamma functions \cite{Ritbergen1997,Chetyrkin1997}, one finds that $\beta_0$,
$\beta_1$, $\gamma_0$, and $\gamma_1$ are not changed and thus universal \cite{Caswel1974},
while modifications are necessary for $\beta_{j\ge 2}$ and $\gamma_{j\ge 2}$.

The exact solutions of Eqs.~(\ref{alphadif}) and (\ref{msdif})
can be obtained by separation of variables, as shown in the appendix A.
At one-loop level, the running coupling and running quark mass of strange quarks
are, respectively, given by
\begin{equation} \label{alphamN1}
\alpha(u)=\frac{1}{\beta_0\ln(u^2/\Lambda^2)}, \ \
m_s=\hat{m}_s\alpha^{\gamma_0/\beta_0},
\end{equation}
where $\beta_0=11/4-N_\mathrm{f}/6$, $\gamma_0=1$,
 $\Lambda$ and $\hat{m}_s$ are the QCD scale parameters
 respectively for the coupling and strange quark mass.

With the requirement of continuity at the threshold of
heavy quark masses and the initial condition
$\alpha_s(M_\mathrm{Z}) = 0.1185$
(where $M_\mathrm{Z}$=91.1876 MeV is the mass of Z bosons)
and $m_s(2 \mbox{GeV}) = 93.5$ MeV \cite{Olive2014}, one can get
distinct coupling scale $\Lambda_{N_\mathrm{f}}$ and mass scale $\hat{m}_{s,N_\mathrm{f}}$ for different effective number of flavors, i.e. $\Lambda_{3-6}$ and $\hat{m}_{s,3-6}$,
respectively corresponding to $u$ in different range, i.e.,
$u < m_c$, $m_c < u < m_b$, $m_b < u < m_t$, $u>m_t$,
where $m_c = 1.275$ GeV, $m_b = 4.18$ GeV, $m_t = 173.21$ GeV \cite{Olive2014}.
The results to four-loop level and for different number of
flavors are given in TABLE I. Because we are considering
three flavors to the leading order, in the following calculations
we take $\Lambda = \Lambda_3 = 146$ MeV and $\hat{m}_s = \hat{m}_{s,3} = 280$
MeV, respectively.


\begin{table}
\caption{The QCD scale parameters $\Lambda_{N_\mathrm{f}}$ and $\hat{m}_{s,N_\mathrm{f}}$ in MeV
to four-loop level for the number of flavors $N_\mathrm{f}$ from 3 to 6.}
\begin{tabular}{c|cccccccc}\hline
loop No.\ & $\Lambda_3$ & $\Lambda_4$ & $\Lambda_5$ & $\Lambda_6$ & $ m_{s,3}$ & $ m_{s,4}$ & $ m_{s,5}$ & $ m_{s,6}$ \\ \hline
1 & 146.2 &122.9 & 90.44 & 44.03 & 279.9 & 303.5 & 339.5 & 401.4 \\
2 & 365.3 &309.5 & 217.9 & 90.65 & 248.2 & 263.8 & 291.0 & 341.4 \\
3 & 342.4 &297.1 & 213.4 & 89.93 & 242.5 & 257.0 & 283.4 & 332.4 \\
4 & 339.1 &295.7 & 212.7 & 89.67 & 240.3 & 254.3 & 280.4 & 329.0 \\ \hline
\end{tabular}
\end{table}

Now we need to choose the relation between the renamalization point $u$
and the chemical potentials. In principle, the choice is not unique.
In Ref.\ \cite{Fraga2005}, it is chosen to be
\begin{equation}\label{u1}
u=\frac{2}{3}(\mu_\mathrm{u}+\mu_\mathrm{d}+\mu_\mathrm{s}).
\end{equation}

To maintain weak equilibrium, the chemical potentials satisfy
\begin{equation}\label{chemical_equilibrium}
\mu_\mathrm{u}+\mu_\mathrm{e}=\mu_\mathrm{d}=\mu_\mathrm{s}.
\end{equation}
Furthermore, SQM should be in the state of charge neutrality, i.e.,
\begin{equation}\label{charge_neutrality}
\frac{2}{3}n_\mathrm{u}-\frac{1}{3}n_\mathrm{d}-\frac{1}{3}n_\mathrm{s}-n_\mathrm{e}=0.
\end{equation}
Another condition for the quark system is the baryon number
conservation, reading
\begin{equation}\label{baryon_conservation}
n_\mathrm{b}=\frac{1}{3}(n_\mathrm{u}+n_\mathrm{d}+n_\mathrm{s}).
\end{equation}
For a given baryon number density $n_\mathrm{b}$
we can solve Eqs.\  (\ref{chemical_equilibrium})-(\ref{baryon_conservation})
to obtain all the relevant chemical potentials.
Then the pressure $P$ and the energy density $E$ are given by
\begin{eqnarray}\label{P}
P&=&-\Omega,\\
\label{E}
E&=&\Omega+\sum_i\mu_in_i,
\end{eqnarray}
where $i$ goes over all flavors and electron.

\begin{figure}
  \includegraphics[width=0.47\textwidth]{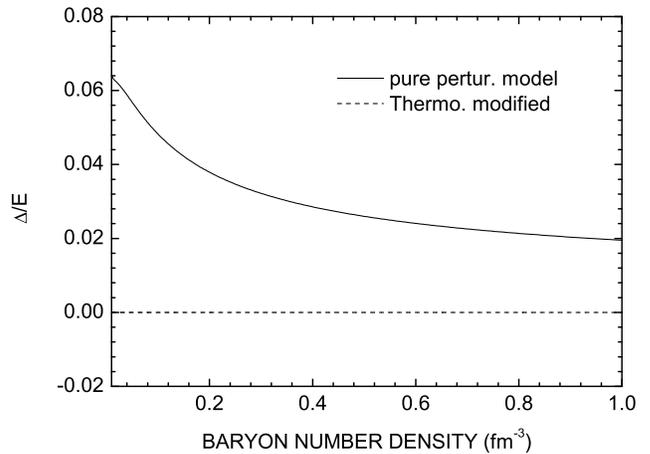}\\
  \caption{The discriminant ratio $\Delta/E$ as a function of density.
  The solid cure is from the pure purtabation model, while the near horizontal line
  is from the EPQ model in the present paper.
  }\label{unmodifiedEperB}
\end{figure}

To check thermodynamic consistency of phenomenological models,
a discriminant $\Delta$, as a function of the baryon number density,
was introduced in Ref.\ \cite{Xia2014} as
\begin{eqnarray}\label{Delta}\nonumber
\Delta&=&P-n_\mathrm{b}^2\frac{\mathrm{d}}{\mathrm{d}n_\mathrm{b}}
\left(\frac{E}{n_\mathrm{b}}\right)\\
&=&P+E-n_\mathrm{b}\frac{\mathrm{d}E}{\mathrm{d}n_\mathrm{b}},
\end{eqnarray}
where $E$ and $P$ are respectively model-given energy density and pressure.
For any thermodynamically consistent models, the discriminant vanishes at arbitrary density.

In order to show the inconsistency degree of the above described pure perturbation model,
we show in FIG.\ \ref{unmodifiedEperB} the density behavior
of the ratio $\Delta/E$ as a function of the density $n_\mathrm{b}$.
From this figure, one can easily find that the discriminant ratio decreases monotonically with increasing density.
This is a clear demonstration that the thermodynamics at higher density
nearly consistent while the thermodynamic inconsistency becomes more and more
serious with decreasing density.

\begin{figure}
  \includegraphics[width=0.48\textwidth]{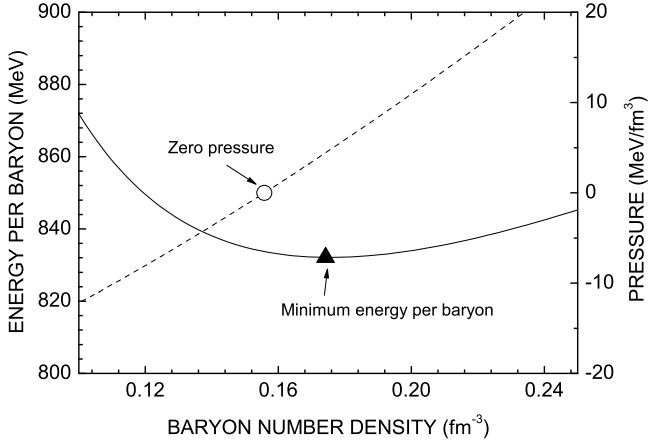}\\
  \caption{Density behavior of the energy per baryon in pQCD
  with a running coupling in Eq.\ (\ref{u1}) and
  a bag constant of $B_0=(135\ \mbox{MeV})^4$.}
\label{unmodifiedEperBwithB120}
\end{figure}

One might think that this problem can be solved by adding a pure constant $B_0$ to
the energy density (subtracting from the pressure) and interpreting it as the vacuum energy density,
just like that had been done in the original bag model.
In FIG.\ \ref{unmodifiedEperBwithB120},
we plot the energy per baryon (left axis) and pressure (right axis)
as function of the density for $\sqrt[4]{B_0}=135$ MeV.
It is obvious in this case that there exists a
minimum energy per baryon (marked with a solid triagle)
and zero pressure (the small open circle). As emphasized
in Ref.~\cite{Peng2000b}, also directly seen from the
first equality of Eq.~(\ref{Delta}), these
two points should appear exactly at the same density.
However, FIG.\ \ref{unmodifiedEperBwithB120} clearly shows that
they are obviously deviate from each, contradicting the
fundamental thermodynamics. This is understandable from the
second equality of Eq.~(\ref{Delta}): adding a constant to $E$ and subtracting it simultaneously
from the pressure $P$ do not influence the value of the discriminant $\Delta$.
In fact, if one draws the density behavior of $\Delta$, one
will find that it is totally overlapped with the case without the constant.

In the following section we will find an additional term that
depends on chemical potentials.
The chemical-potential-dependent term can be neglected at high density
but it takes an important role at lower density, solving
the thermodynamic inconsistency very nicely.
The key point is that the renormalization subtraction $u$
as a function of the chemical potentials can not be arbitrarily taken, such as that in
Eq.~(\ref{u1}). Instead, we choose it to satisfy an equation obtained by the requirement of
thermodynamic consistency.

\section{Extrapolation with thermodynamic consistency}
\label{newTreat}

The thermodynamic potential density of a cold quark system from perturbative
QCD to order $N$ can be generally written as
\begin{equation} \label{Omegapt}
\Omega^{\mathrm{pt}}_N
=\sum_{i=0}^N\omega_i(\mu_\mathrm{u},\mu_\mathrm{d},\mu_\mathrm{s},m_\mathrm{s},\ln\alpha,u)\alpha^i,
\end{equation}
where $\alpha=\alpha_s/\pi=g^2/(4\pi^2)$ is the QCD running coupling,
$m_\mathrm{s}$ is the running mass of a strange quark.
They satisfy the renormalization group
equation given in Eq. (\ref{alphadif}) and (\ref{msdif}).

The corresponding number density for quark flavor q$=$u, d, s can be easily
obtained by the normal thermodynamic relation
$n_\mathrm{q}=-\mathrm{d}\Omega^{\mathrm{pt}}/\mathrm{d}\mu_\mathrm{q}$. Also at the order $N$, it gives
\begin{equation} \label{ni}
n_\mathrm{q}
=\sum_{k=0}^N
 \left[
  -\frac{\partial\omega_k}{\partial\mu_\mathrm{q}}
  -\frac{\partial\omega_k}{\partial{u}}\frac{\partial{u}}{\partial\mu_\mathrm{q}}
  +\frac{2}{u}\frac{\partial{u}}{\partial\mu_q}
   \sum_{i=0}^{k-1}f_{i,k-i-1}
 \right]\alpha^k,
\end{equation}
where $f_{i,j}$ ($j=k-i-1$) is zero if $i<0$ or $j<0$, otherwise it is defined to be
\begin{equation}
f_{i,j}\label{f_i,j}
=\left(i\omega_i+\frac{\partial\omega_i}{\partial\ln\alpha}\right)\beta_j
+m_\mathrm{s}\frac{\partial\omega_i}{\partial m_\mathrm{s}}\gamma_j.
\end{equation}

We write the whole thermodynamic potential density of the system as
\begin{equation} \label{Omegatot}
\Omega
=\Omega_N^{\mathrm{pt}}+\Omega^\prime.
\end{equation}
Here $\Omega_\mathrm{N}^{\mathrm{pt}}$ is the perturbative contribution, while
$\Omega'$ is the non-perturbative contribution. To determine $\Omega'$, we
require that it makes $\Omega$ satisfy the fundamental thermodynamic equation
\begin{equation}
\mbox{d}\Omega
=-S\mbox{d}T-\sum_in_i\,\mbox{d}\mu_i,
\end{equation}
where $S$ is the entropy density at temperature $T$. At zero temperature
the first term vanishes and we have $\mbox{d}\Omega=-\sum_in_i\,\mbox{d}\mu_i$.
Substituting Eqs.~(\ref{Omegatot}), (\ref{Omegapt}), and (\ref{ni})
into this equality, we immediately obtain
\begin{equation}
\mbox{d}\Omega^\prime
=\frac{\partial\Omega^\prime}{\partial\mu_\mathrm{u}}\mbox{d}\mu_\mathrm{u}
+\frac{\partial\Omega^\prime}{\partial\mu_\mathrm{d}}\mbox{d}\mu_\mathrm{d}
+\frac{\partial\Omega^\prime}{\partial\mu_\mathrm{s}}\mbox{d}\mu_\mathrm{s},
\end{equation}
where the partial derivatives are
\begin{eqnarray}
\label{pOppmuu}
\frac{\partial\Omega^\prime}{\partial\mu_\mathrm{u}}
&=&
\frac{2}{u}\frac{\partial{u}}{\partial\mu_\mathrm{u}}
\sum_{k,i}^{(N,{\cal N})}f_{i,k-i-1}\alpha^k\equiv G_1(\mu_\mathrm{u},\mu_\mathrm{d},\mu_\mathrm{s}), \\
\label{pOppmud}
\frac{\partial\Omega^\prime}{\partial\mu_\mathrm{d}}
&=&
 \frac{2}{u}\frac{\partial{u}}{\partial\mu_\mathrm{d}}
 \sum_{k,i}^{(N,{\cal N})}f_{i,k-i-1}\alpha^k\equiv G_2(\mu_\mathrm{u},\mu_\mathrm{d},\mu_\mathrm{s}), \\
\label{pOppmus}
\frac{\partial\Omega^\prime}{\partial\mu_\mathrm{s}}
&=&
 \frac{2}{u}\frac{\partial{u}}{\partial\mu_\mathrm{s}}
 \sum_{k,i}^{(N,{\cal N})}f_{i,k-i-1}\alpha^k\equiv G_3(\mu_\mathrm{u},\mu_\mathrm{d},\mu_\mathrm{s}).
\end{eqnarray}
The double summation in Eqs. (\ref{pOppmuu})-(\ref{pOppmus}) is given by
\begin{equation}
\sum_{k,i}^{(N,{\cal N})}
\equiv\sum_{k=N+1}^{N+{\cal N}}\sum_{i=\mathrm{max}(0,k-{\cal N})}^{\mathrm{min}(k-1,N)},
\end{equation}
Therefore, the additional term $\Omega^\prime$ is given by a path integral as
\begin{equation} \label{Omegap01}
\Omega^\prime
=\int_{\mu_0}^{\mu}
(
 G_1\mbox{d}\mu_\mathrm{u}
+G_2\mbox{d}\mu_\mathrm{d}
+G_3\mbox{d}\mu_\mathrm{s}
)+B_0.
\end{equation}
where $\mu_0=(\mu_\mathrm{u0},\mu_\mathrm{d0},\mu_\mathrm{s0})$ is a starting
point for the path integral which is fixed to be
$\mu_\mathrm{u0}=\mu_\mathrm{d0}=\mu_\mathrm{s0}=313$ MeV in the present calculation,
while its moving effect is boiled down to another constant $B_0$.

As everyone knows, the thermodynamic potential is a state function.
Therefore, $\Omega^\prime$\ should be independent of the path,
namely, the integration in Eq.~(\ref{Omegap01}) is path-independent.
This requires that the integrands satisfy the Cauchy conditions
\begin{equation}\label{Cauchy_condition}
\frac{\partial G_1}{\partial\mu_\mathrm{d}}=\frac{\partial G_2}{\partial\mu_\mathrm{u}},\ \ \
\frac{\partial G_2}{\partial\mu_\mathrm{s}}=\frac{\partial G_3}{\partial\mu_\mathrm{d}},\ \ \
\frac{\partial G_1}{\partial\mu_\mathrm{s}}=\frac{\partial G_3}{\partial\mu_\mathrm{u}}.
\end{equation}
It is easy to prove that only two of them are independent.

In the present paper, let's take the first-order of $\Omega_N^{\mathrm{pt}}$ as an example
, i.e. $\Omega_1^{\mathrm{pt}}$, with running coupling $\alpha$ and running mass $m_\mathrm{s}$
expanded to first order which are explicitly given by Eq. (\ref{alphamN1}).
In this case, we have
\begin{eqnarray}
\label{G123}
\nonumber
G_1&=&\frac{2}{u}\frac{\partial u}{\partial\mu_\mathrm{u}}f_{1,0}\alpha^2,\ \ \ \ \
G_2=\frac{2}{u}\frac{\partial u}{\partial\mu_\mathrm{d}}f_{1,0}\alpha^2,\\
G_3&=&\frac{2}{u}\frac{\partial u}{\partial\mu_\mathrm{s}}f_{1,0}\alpha^2.
\end{eqnarray}
With the notation $\tau\equiv\sqrt[4]{\mu_\mathrm{u}^4
+\mu_\mathrm{d}^4+\mu_\mathrm{s}^4}$
and the expressions of $\omega_0$ and $\omega_1$, i.e.
\begin{eqnarray}\nonumber
\omega_0&=&\frac{-1}{4\pi^2}\left[\tau^4
 +\mu_\mathrm{s}\nu_\mathrm{s}\big(\mu_\mathrm{s}^2-\frac{5}{2}m_\mathrm{s}^2\big)
 +\frac{3}{2}m_\mathrm{s}^4\mathrm{ach}\frac{\mu_\mathrm{s}}{m_\mathrm{s}}
\right],\\
\omega_1
&=&
 \frac{1}{2\pi^2}\Bigg[
  \tau^4-2\nu_s^4+3\left(\mu_s\nu_s-m_s^2\mathrm{ach}\frac{\mu_\mathrm{s}}{m_\mathrm{s}}\right)^2
\nonumber\\
&& \hspace{-3mm}
+m_s^2 \Big(6\ln\frac{u}{m_s}+4 \Big)
 \bigg(\mu_s\nu_s-m_s^2\mathrm{ach}\frac{\mu_\mathrm{s}}{m_\mathrm{s}}\bigg) \Bigg],
\end{eqnarray}
one can derive the explicit expression of $\ f_{1,0}$, giving
\begin{eqnarray}\label{f10}\nonumber
f_{1,0}&=&\beta_0\omega_1+\gamma_0m_\mathrm{s}\frac{\partial\omega_1}{\partial m_\mathrm{s}}\\
&=&
\frac{9}{8\pi^2}\left(\mu_\mathrm{u}^4+\mu_\mathrm{d}^4+\mu_\mathrm{s}^4\right)
+\frac{75m_\mathrm{s}^4}{8\pi^2}\mbox{ach}^2\left(\frac{\mu_\mathrm{s}}{m_\mathrm{s}}\right)
\nonumber\\
&&
+\frac{m_\mathrm{s}^2}{8\pi^2}
\Bigg[41\mu_\mathrm{s}^2-50m_\mathrm{s}^2+44\mu_\mathrm{s}\nu_\mathrm{s}
\nonumber \\
&&
+6\ln\frac{u}{m_\mathrm{s}}
\Bigg(17\mu_\mathrm{s} \nu_\mathrm{s}
-25m_\mathrm{s}^2 \mbox{ach}\frac{\mu_\mathrm{s}}{m_\mathrm{s}}\Bigg)
\nonumber \\
&&
-2\mbox{ach}\frac{\mu_\mathrm{s}}{m_\mathrm{s}}\left(
38m_\mathrm{s}^2+51\mu_\mathrm{s}\nu_\mathrm{s}
\right)
\Bigg],
\end{eqnarray}

Substituting Eq. (\ref{G123}) into the first two Cauchy conditions, we have
\begin{eqnarray}\label{Cauchy_conditionp}
\frac{\partial u}{\partial\mu_\mathrm{u}}\frac{\partial f_{1,0}}{\partial\mu_\mathrm{d}}
=\frac{\partial u}{\partial\mu_\mathrm{d}}\frac{\partial f_{1,0}}{\partial\mu_\mathrm{u}},\
\frac{\partial u}{\partial\mu_\mathrm{d}}\frac{\partial f_{1,0}}{\partial\mu_\mathrm{s}}
=\frac{\partial u}{\partial\mu_\mathrm{s}}\frac{\partial f_{1,0}}{\partial\mu_\mathrm{d}}.
\end{eqnarray}
From Eq. (\ref{f10}), one can get the partial derivatives of $f_{1,0}$
with respect to $\mu_\mathrm{u}$ and $\mu_\mathrm{d}$ respectively, i.e.
\begin{equation}\label{pfpmuumud}
\frac{\partial f_{1,0}}{\partial\mu_\mathrm{u}}=\frac{9\mu_\mathrm{u}^3}{2\pi^2},\
\frac{\partial f_{1,0}}{\partial\mu_\mathrm{d}}=\frac{9\mu_\mathrm{d}^3}{2\pi^2}.
\end{equation}
Using Eq. (\ref{pfpmuumud}), the first equation in Eq. (\ref{Cauchy_conditionp}) becomes
\begin{equation}
\mu_\mathrm{u}^3\frac{\partial u}{\partial \mu_\mathrm{d}}=
\mu_\mathrm{d}^3\frac{\partial u}{\partial \mu_\mathrm{u}}.
\end{equation}
This equation means that the solution of $u$ is a function of $\mu_\mathrm{s}$ and
$\rho\equiv\sqrt[4]{\mu_\mathrm{u}^4+\mu_\mathrm{d}^4}$, i.e.
$u=u(\rho,\mu_\mathrm{s})$. However, this
function is not necessarily explicit, and can be generally
implicit. So we assume it is determined by the following implicit equation
\begin{equation}
\Phi(\rho,\mu_\mathrm{s},u)=0.
\end{equation}

In order to find the form of $\Phi$, we give the partial derivatives of
$u$ with respect to $\mu_\mathrm{d}$ and $\mu_\mathrm{s}$, i.e.
\begin{equation}\label{umud_umus}
\frac{\partial u}{\partial\mu_\mathrm{d}}
=\frac{\partial u}{\partial\rho}\frac{\partial\rho}{\partial\mu_\mathrm{d}}
=-\frac{\Phi_\rho}{\Phi_u}\frac{\partial\rho}{\partial\mu_\mathrm{d}},\
\frac{\partial u}{\partial\mu_\mathrm{s}}=
-\frac{\Phi_{\mu_\mathrm{s}}}{\Phi_u},
\end{equation}
where the notations
$\Phi_x\equiv {\partial\Phi}/{\partial x}$\ $(x=\rho, u, \mu_\mathrm{s})$
have been used.
Then substituting Eq. (\ref{umud_umus}) to the second equation in Eq. (\ref{Cauchy_conditionp}) leads to
\begin{equation}\label{umud_umusp}
\frac{8\pi^2}{9}\frac{\partial f_{1,0}}{\partial\mu_\mathrm{s}}\frac{\partial\Phi}{\partial\rho}=
4\rho^3\frac{\partial\Phi}{\partial\mu_\mathrm{s}},
\end{equation}
where $\frac{\partial f_{1,0}}{\partial\mu_\mathrm{s}}$ can be obtained from Eq. (\ref{f10}), i.e.,
\begin{eqnarray}\label{pfpmus}
\frac{\partial f_{1,0}}{\partial\mu_\mathrm{s}}
&=&
\frac{m_\mathrm{s}^2}{2\pi^2\nu_\mathrm{s}^2}
 \left[
  \nu_\mathrm{s}(22\mu_\mathrm{s}^2-30m_\mathrm{s}^2)
  +\mu_\mathrm{s}(5m_\mathrm{s}^2-14\mu_\mathrm{s}^2)
 \right]
\nonumber\\
&+&
 \frac{9\mu_\mathrm{s}^5}{2\pi^2\nu_\mathrm{s}^2}
 -\frac{3m_\mathrm{s}^2}{2\pi^2\nu_\mathrm{s}}(17\mu_\mathrm{s}^2-21m_\mathrm{s}^2)
\mbox{ln}\frac{\mu_\mathrm{s}+\nu_\mathrm{s}}{u}.
\end{eqnarray}

For Eq. (\ref{umud_umusp}) to be fulfilled, we choose
\begin{equation} \label{PhirhoPhiMus}
\frac{\partial \Phi}{\partial\rho}=4\rho^3, \ \
\frac{\partial\Phi}{\partial\mu_\mathrm{s}}
=\frac{8\pi^2}{9}\frac{\partial f_{1,0}}{\partial\mu_\mathrm{s}}.
\end{equation}

Solving the equalities in Eq.~(\ref{PhirhoPhiMus}), we can find the solution
\begin{equation} \label{Lgen}
\Phi=\frac{8\pi^2}{9}f_{1,0}-\phi(u),
\end{equation}
where $\phi(u)$ is an arbitrary function of $u$.
In fact, integrating the first equality in Eq.~(\ref{PhirhoPhiMus}), one can
get the solution as $\Phi(\rho,\mu_\mathrm{s},u)=\rho^4+\varphi(\mu_\mathrm{s},u)$,
where the integration constant $\varphi$ (with respect to the
variables $\mu_\mathrm{s}$ and $u$) can be obtained by substituting into the second equality of Eq.~(\ref{PhirhoPhiMus}) as
$$
\varphi(\mu_\mathrm{s},u)
=\frac{8\pi^2}{9}\int \frac{\partial f_{1,0}}{\partial\mu_\mathrm{s}}\mbox{d}\mu_\mathrm{s}
-\rho^4-\phi(u),
$$
which gives the solution in Eq.~(\ref{Lgen}) immediately.

For convenience and simplicity, we have chosen the function
$\phi(u)=N_{\mathrm{f}}(u/C)^4$ with constant $C$ being a model
parameter which can be fixed in a reasonable region by the
common knowledge of modern nuclear physics.
So, from the solution of Eq. (\ref{Cauchy_conditionp}),
one can find that to solve the problem of
inconsistency in thermodynamics, the relation between the renormalization
subtraction point $u$ and chemical potentials should be
obtained by solving the following equation
\begin{equation}\label{mu_u}
\frac{8\pi^2}{9}f_{1,0}(\rho,\mu_\mathrm{s},u)-\frac{N_{\mathrm{f}}}{C^4}u^4=0.
\end{equation}

As is well known, the mass effect from strange quarks can be
ignored at very high density. In this case the relation
between the renormalization subtraction point and chemical potentials can be given in a
simple form, i.e.,
\begin{equation}
u=C\sqrt[4]{\frac{\mu_u^4+\mu_d^4+\mu_s^4}{N_{\mathrm{f}}}},
\end{equation}
which can be obtained by taking $m_s\rightarrow\ 0$ in Eq. (\ref{mu_u}).

In order to get the EOS, in general one should numerically solve
Eq. (\ref{mu_u}). Then substituting Eq. (\ref{G123}) to
Eq. (\ref{Omegap01}) and numerically integrating it, one can get the
quantity $\Omega'$ which is essential in fixing the problem of
thermodynamic inconsistency. And the partial derivatives of renormalization subtraction
point $u$ with respect to chemical potentials $\mu_\mathrm{q}\
(\mathrm{q}=\mathrm{u}, \mathrm{d}, \mathrm{s})$ in Eq. (\ref{G123})
are
\begin{equation}\label{pupmuq}
\frac{\partial u}{\partial\mu_\mathrm{q}}
=\frac{ \partial f_{1,0}/\partial\mu_\mathrm{q} }
      { \frac{9N_\mathrm{f}u^3}{2\pi^2C^4}
       -\frac{\partial f_{1,0}}{\partial u}
       -\frac{\partial f_{1,0}}{\partial m_\mathrm{s}}
       \frac{\partial m_\mathrm{s}}{\partial\alpha}
       \frac{\partial\alpha}{\partial u}
       },
\end{equation}
where $\partial f_{1,0}/\partial\mu_\mathrm{q}$ are already given in Eqs. (\ref{pfpmuumud})
and (\ref{pfpmus}). Other relevant derivatives of $f_{1,0}$ in Eq. (\ref{pupmuq}) are, respectively
\begin{equation}
\frac{\partial f_{1,0}}{\partial u}
=\frac{3m_\mathrm{s}^2}{4\pi^2 u}
\left(
 17\mu_\mathrm{s}\nu_\mathrm{s}-25m_\mathrm{s}^2\mbox{ach}
 \frac{\mu_\mathrm{s}}{m_\mathrm{s}}
\right)
\end{equation}
and
\begin{eqnarray}
\nonumber
\frac{\partial f_{10}}{\partial m_\mathrm{s}}&=&
\frac{75}{2\pi^2}m^3\mathrm{ach}^2\left(\frac{\mu_\mathrm{s}}{m_\mathrm{s}}\right)
-\bigg[\frac{75}{\pi^2}m_\mathrm{s}^3\ln\frac{u}{m_\mathrm{s}}
\\
\nonumber
&&+\frac{m_\mathrm{s}}{4\pi^2\nu_\mathrm{s}}
(78\mu_\mathrm{s}\nu_\mathrm{s}^2+77m_\mathrm{s}^2\nu_\mathrm{s}
+24\mu_\mathrm{s}^3)\bigg]
\\
\nonumber
&&\times\mathrm{ach}\frac{\mu_\mathrm{s}}{m_\mathrm{s}}
-\frac{3m_\mathrm{s}\mu_\mathrm{s}}{2\pi^2\nu_\mathrm{s}}
(13m_\mathrm{s}^2-17\mu_\mathrm{s}^2)\ln\frac{u}{m_\mathrm{s}}
\\
\nonumber
&&
+\frac{m_\mathrm{s}}{4\pi^2\nu_\mathrm{s}^2}(23m_\mathrm{s}^2\mu_\mathrm{s}\nu_\mathrm{s}
+92\mu_\mathrm{s}^4+100m_\mathrm{s}^4\\
&&-7\mu_\mathrm{s}^3\nu_\mathrm{s}-192m_\mathrm{s}^2\mu_\mathrm{s}^2).
\end{eqnarray}
The remaining derivatives on the right-hand side of Eq. (\ref{pupmuq}) are
\begin{eqnarray}
\frac{\partial m_\mathrm{s}}{\partial\alpha}
=\frac{4\hat{m_\mathrm{s}}}{9}\alpha^{-\frac{5}{9}}\
\mathrm{and}\
\frac{\partial\alpha}{\partial u}=-\frac{9}{2}\frac{\alpha^2}{u}.
\end{eqnarray}

\section{EOS of strange quark matter}
\label{EOS-SQM}

As usually done, we assume SQM to be a mixture of interacting quarks and free electrons.
So the total thermodynamic potential density $\Omega_\mathrm{tot}$ reads
\begin{equation}\label{Omega_tot}
\Omega_{\mathrm{tot}}
=\Omega^{\mathrm{pt}}-\frac{\mu_e^4}{12\pi^2}+\Omega^\prime,
\end{equation}
where the first term is the perturbative contribution in Eq.~(\ref{Omegapt}) whose
concrete form is given to leading order in Eqs.~(\ref{Omegaud}) and (\ref{Omegas}).
The second term is the contribution from electrons
treated as free particles because they do not participate in the
strong interactions, and the last term is given in Eq.~(\ref{Omegap01}), determined
by thermodynamic consistency requirement to consider non-perturbative effect.

For a given baryon number density, one can solve
Eqs. (\ref{chemical_equilibrium})-(\ref{baryon_conservation})
with the help of Eqs.~(\ref{nu-nd}) and (\ref{ns}) to obtain the relevant
particle chemical potentials, then all other quantities can be thermodynamically
obtained. But in the present model, there are still two parameters, $B_0$ and $C$,
to be determined by stability arguments.

\begin{figure}[thb]
\centering
\includegraphics[width=0.48\textwidth]{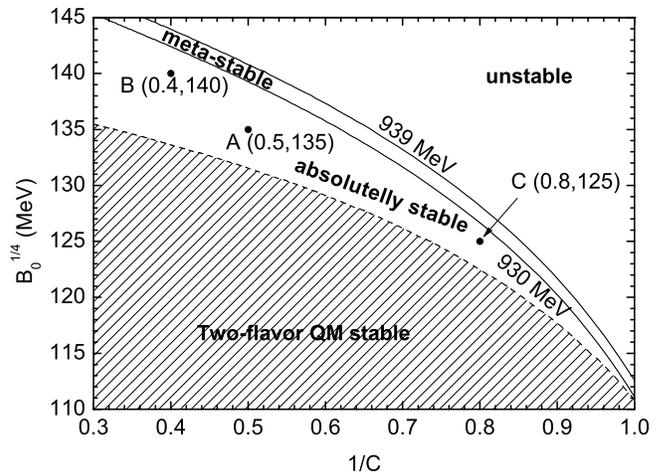}
\caption{
The parameter range in $B_0^{1/4}$ vs $1/C$ plane.
The shaded region is forbidden where
two-flavor quark matter is stable.
The energy per baryon of three-flavor quark matter
is bigger than 939 MeV in the right upper region marked with `unstable',
less than 939 MeV but bigger than 930 MeV in the region with `metastable',
smaller than 930 MeV in the absolutely stable region
where three sets of parameters A, B, and C are indicated by solid dots.}
\label{BC}
\end{figure}

As is well known, the energy per baryon of two-flavor quark matter should be
bigger than $930$ MeV ($E/n_\mathrm{b}>930$ MeV), in order not to contradict
standard nuclear physics. Therefore, the shaded region in FIG.\ \ref{BC} is
forbidden. If the energy per baryon of three flavor quark matter is less than $930$ MeV,
the SQM is absolutely stable; if it is bigger than $930$ MeV, but smaller than $939$ MeV,
the SQM is metastable; otherwise, the SQM is untable. These different regions are indicated
in FIG.\ \ref{BC}.

To investigate the properties of SQM in the present EPQ model,
we choose three typical sets of parameters in the absolutely stable region,
i.e., $(1/C,\sqrt[4]{B_0}/\mbox{MeV})=(0.5, 135), (0.4,140), (0.8,125)$,
respectively represented with capital letters A, B, and C in FIG.\ \ref{BC}.

\begin{figure}
  \includegraphics[width=0.48\textwidth]{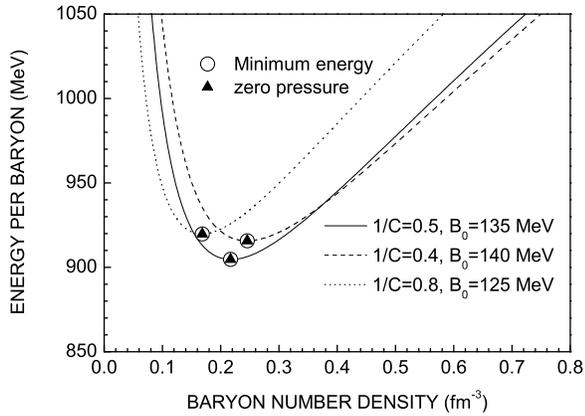}\\
  \caption{Density behavior of the energy per baryon.
  It is obvious that the minimum energy (the triangle) for each curve
  locates exactly at the density
  corresponding to zero pressure (the circle). }
\label{EperB}
\end{figure}

\begin{figure}
  \includegraphics[width=0.48\textwidth]{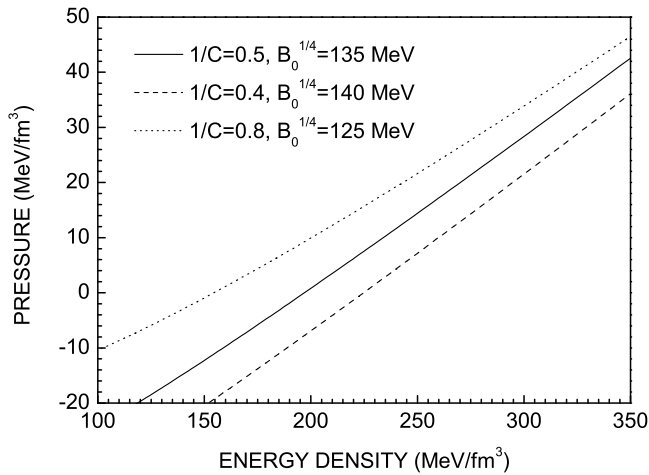}\\
  \caption{The EOS of SQM for the three
  typical parameter sets. }
\label{EOS}
\end{figure}

Let us first check the consistency of EPQ model.
FIG.\ \ref{EperB} shows the density behavior of the energy per baryon with different parameters.
It is obvious that the minimum energy per baryon for each curve locates exactly at the density
corresponding to zero pressure. This consistency in thermodynamics
can be seen in FIG.\ \ref{unmodifiedEperB} where the value of $\Delta/E$
for the present model is zero at all density.
In addition, the three minimum points in this figure correspond to
absolutely stable SQM, as expected. It is also found that the minimum energy per baryon in fact becomes
bigger with decreasing $C$ and/or increasing $B_0$.

\begin{figure}
  \includegraphics[width=0.48\textwidth]{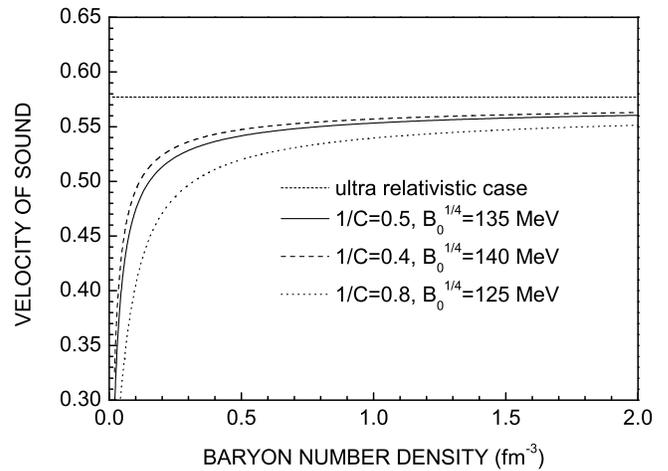}\\
  \caption{
  Velocity of sound in SQM. The nearly horizontal line is for
  the ultra relativistic case,
  the three lower cures are from the present
  model for the three typical sets of parameters.
  }
\label{VoS}
\end{figure}

FIG.\ \ref{EOS} gives the EOS of SQM.
The stiffness of EOS obviously varies with the model parameters
$C$ and $B_0$, e.g., EOS would become stiffer with bigger $C$ and/or smaller $B_0$.
In the next section we will see that this means a bigger
maximum mass of compact stars (see FIG.\ \ref{mr}).

To check the impact of the additional term $\Omega'$,
we plot, in FIG.\ \ref{VoS}, the density behavior of the
velocity of sound calculated by
\begin{equation}
v=\sqrt{\Big|\frac{\mathrm{d}P}{\mathrm{d}E}\Big|}.
\end{equation}
We have noted that the density behavior of sound velocity
is greatly affected by the stiffness of EOS, and it is
understandable that stiffer EOS corresponds to fast velocity of sound.
At very high density, however, they all approach to the ultra relativistic case.
In addition, we would like to point out that the
sound velocity is independent of the parameter $B_0$.

\begin{figure}
  \includegraphics[width=0.48\textwidth]{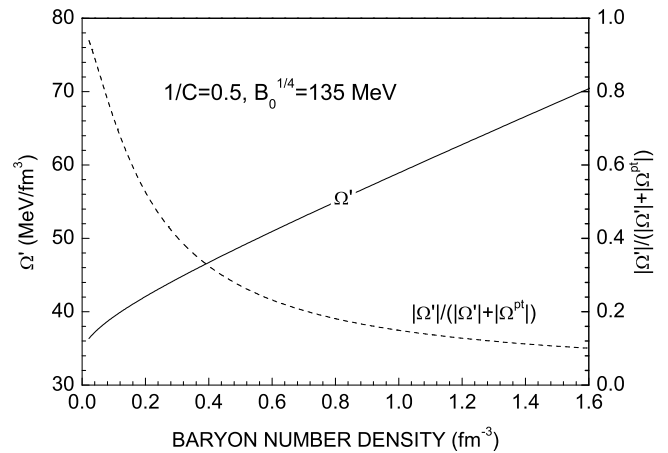}\\
  \caption{The density behavior of $\Omega'$ and its relative importance.}
\label{Omegap}
\end{figure}

FIG.\ \ref{Omegap} shows the density behavior of $\Omega'$ and its relative importance.
From this figure one can see that
although $\Omega'$ increases with increasing density,
the relative importance decreases with increasing density.
That means $\Omega'$ plays a relatively important
role at lower density while it is ignorable at high density.
In this regard it is similar to a
chemical-potential-dependent bag constant.

\section{Structure of strange stars}
\label{struc-stars}

Neutron stars are the main targets of future observatories \cite{Ray2012,Feroci2012,Watts2014,Arzoumanian2014}
and have long been interesting objects of many theoretical and
observational investigations
\cite{Itoh1970,Alcock1986,Haensel1986,Prakash1997,
Antoniadis2013,Demorest2010,Roupas2015}.
Because their inner matter is very dense, they become the most
promising places to find quark matter \cite{Heiselberg2000,Buballa2014}.

In general case, such a compact object may be a hybrid star
with pure quark core and hadronic crust \cite{Li2015}.
Because SQM can be self-bound, i.e., its internal pressure
can be zero at a definite density (see the minima in Fig. \ref{EperB}),
the whole star can be converted to a pure quark star,
for example, as a strong deflagration process during a
few milliseconds \cite{Herzog2011},
or seeded with slets \cite{Peng2006plb} by the self-annihilating weakly
interacting massive particles \cite{Angeles2010}, etc.


\begin{figure}
  \includegraphics[width=0.45\textwidth]{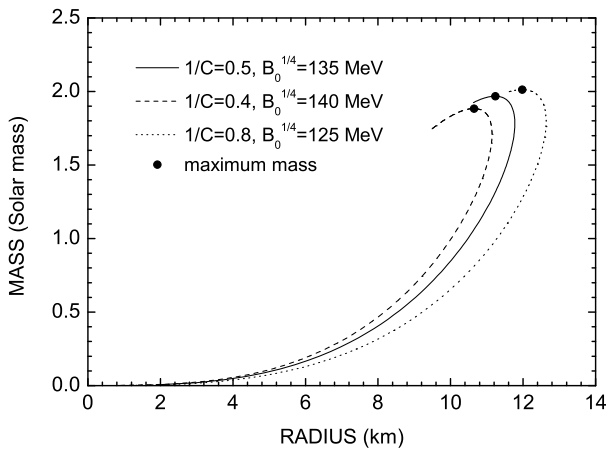}\\
  \caption{The mass-radius relation of quark stars.
}\label{mr}
\end{figure}

In the preceding sections,
we have developed an enhanced version of a perturbative
QCD treatment of the dense quark matter
by thermodynamic consistency requirement.
Now we apply it to study the structure of quark stars.
For this purpose we should solve the Tolman-Oppenheimer-Volkov (TOV)
equation \cite{Lipunov}
\begin{equation}
\frac{\mathrm{d}P}{\mathrm{d}r}
=-\frac{GmE}{r^2}\frac{(1+P/E)(1+4\pi r^3P/m)}{1-2Gm/r},
\end{equation}
with the subsidiary condition
\begin{equation}
\frac{\mathrm{d}m}{\mathrm{d}r}=4\pi r^2E,
\end{equation}
where $G=6.707\times10^{-45}\ \mathrm{MeV}^{-2}$ is the gravitational constant,
$r$ is the distance from the center of a quark star,
and $P$ and $E$ are the pressure and energy density with their mutual relation given by EOS.
One can refer to Ref. \cite{Peng2000b} for a concise process of how to solve this equation.

On application of the equations of state in Fig.\ \ref{EOS},
we can get the mass-radius relation in FIG.\ \ref{mr}
for the typical parameter sets indicated in the legend.
This figure shows several feathers of quark stars.

(1) The radius of a quark star can be in principle
very small, i.e., there is no lower bound to the radius.
This is very different from the normal neutron stars
whose radius are normally greater than a critical value.

(2) For a given set of parameters,
the star radius first increases with increasing
the star mass, until a maximum radius is reached.
After that, the radius decreases until
the maximum star mass is arrived.

(3) The maximum star mass depends on parameters.
It actually increases with increasing $C$, while decreases with increasing $B_0$.
Specially for the typical parameters $C=2$
and $B_0=(135\ \mbox{MeV})^4$,
The maximum mass is about two times the solar mass,
consistent with the recent high-mass observations \cite{Antoniadis2013,Demorest2010}.


\begin{figure}
  \includegraphics[width=0.49\textwidth]{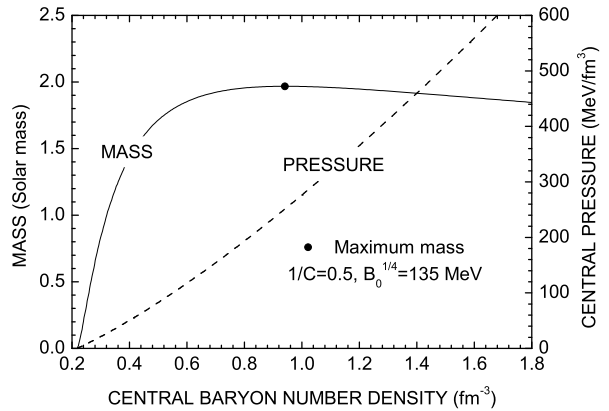}\\
  \caption{The mass and central pressure of the quark star as functions
  of the central density for the typical parameters
  $1/C=0.5$ (or $C=2$) and $B_0^{1/4}=135$ MeV.
  }\label{nbEP}
\end{figure}

\begin{figure}
  \includegraphics[width=0.48\textwidth]{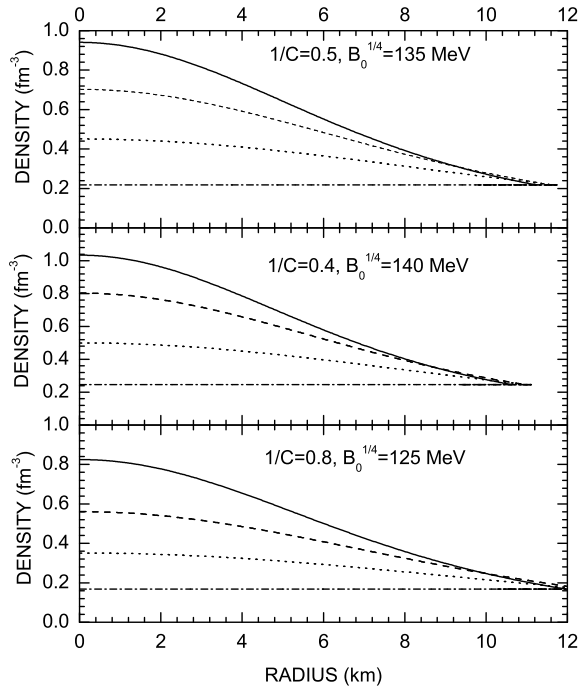}\\
  \caption{
  Density profiles for different sets of parameters
  indicated in the legends. The solid curve in each panel
  is for the highest central density corresponding
  to the quark star with maximum mass,
  while the horizontal represents the surface density.}
\label{Rnb}
\end{figure}

\begin{table}
\centering
\caption{\label{tab2}
Characteristic quantities for the typical parameter sets
$(C^{-1},{B_0^{1/4}}/{\mathrm{MeV}})=(0.5,135)$, (0.4,140), and (0.8,125).
The 2---6th rows give, respectively, the maximum mass $M_\mathrm{max}$,
the radius corresponding to the maximum mass $R(M_\mathrm{max})$,
the highest central density at the maximum mass
$n_\mathrm{max}$,
the surface density $n_0$ and the corresponding energy per baryon $E_0/n_0$.
}
\begin{tabular}{|c|c|c|c|}
  \hline
  $(1/C,B_0^{1/4}/\mathrm{MeV})$  & (0.5,135) & (0.4,140) & (0.8,125) \\ \hline
  $M_\mathrm{max}/M_\odot$       & 1.968    & 1.884     & 2.013 \\ \hline
  $R(M_\mathrm{max})$\, [km]        & 11.2   & 10.6    & 12.0 \\ \hline
  $n_\mathrm{max}$\, [fm$^{-3}]$    & 0.941    & 1.034     & 0.8250 \\ \hline
  $n_0$\, [fm$^{-3}]$               & 0.2177    & 0.2457     & 0.1683 \\ \hline
  $E_0/n_0$\, [MeV]                 & 904.7     & 915.7      & 919.8 \\ \hline
\end{tabular}
\end{table}

To understand the existence of a maximum star mass,
we show, in FIG.\ \ref{nbEP} for the parameter set A,
the star mass as a function of the central density,
with the central pressure also given on the right axis.
One can easily find that the star mass first increases with increasing the central density to a critical density $n_\mathrm{max}$. After this density, the star mass
decreases with increasing the central density,
and the star itself becomes mechanically unstable.
The central pressure is always an
increasing function of the central density.
It approaches to zero if the quark star mass becomes
zero. Please note, the corresponding central density
to zero pressure is nonzero.
Instead, it is a definite value corresponding to the surface density of the quark star.

The density of a quark star is not uniformly distributed.
In FIG.\ \ref{Rnb}, we plot the density profiles with different panels
for different parameters.
For each parameter set, the upmost curve corresponds to
the star with the maximum mass, $r=0$ corresponds to the
central density. At the star surface, the pressure is zero.
This point corresponds to the minimum energy per baryon
in FIG.\ \ref{EperB}. Therefore, the surface density of the
quark star is not zero, i.e., the star
has a sharp surface. In Tab.\ \ref{tab2},
we list some characteristic quantities
of quark stars for each parameter set,
including the maximum star mass to the solar mass $M_\mathrm{max}/M_\odot$,
the corresponding radius $R(M_\mathrm{max})$,
the highest central density $n_\mathrm{max}$, the surface density $n_0$,
and the minimum energy per baryon $E_0/n_0$.

It should be noted that the observation of quark stars
having a sharp surface depends very much on the model
assumption. Especially when one doesn't have a color-flavor
locked phase throughout, the star might have a crust of
ordinary matter supported by electrons extending beyond
the quark surface. Or the outer layers of the quark star
might fragment to strangelets, somewhat similar to a
normal neutron star crust.  Also, for some parameters,
e.g. the case C, the surface density is closer to the
normal nuclear saturation density, which might be an
indication of phase transition to nuclear matter.
To understand possible phase transition in a mixed star,
it is necessary to investigate phase equilibrium
condition in the phase boundary \cite{Peng2008,Li2015}.

\section{summary}
\label{summary}

The perturbative QCD is important to study strongly
interacting matter. However, its naive extension
to comparatively lower density has serious thermodynamic
inconsistency problems due to the running of
the QCD coupling and/or quark masses.
We have tried to extend it by
including an additional term in the thermodynamic potential density.
The additional term is determined by the fundamental
differential equation of thermodynamics.
It takes an important role at
comparatively lower density, but ignorable
at high density, playing the role of a
chemical-potential-dependent bag constant.

On application of the thermodynamically EPQ model,
we study the properties of SQM and structure of quark stars.
It is found that SQM still has the possibility
of being absolutely stable in the present model,
i.e. the minimum energy per baryon could be less than 930 MeV
with zero internal pressure. This leads to
the maximum mass of quark stars as large as
two times the solar mass. The quark stars
in the present model has several features,
such as a sharp surface, no lower bound for the radius,
the maximum mass of about two times the solar mass
and a maximum radius of about 11 kilometers, etc.

Naturally, there several aspects not included
in the present investigations, e.g.
the color superconductivity \cite{Alford1999,Huang2004,Wen2013prd},
a strong external magnetic field \cite{Wen2012,Duncan1992,Ferrer20052006,Dexheimer2014},
etc.
Furthermore, only zero temperature has been considered in the present paper.
To investigate the properties of hot quark matter, such as the quark
gluon plasma produced in high energy heavy ion collisions \cite{HouDF2012,LiuF2013},
the inclusion of finite temperature is of crucial importance.
Therefore, more careful studies are necessary in the future.

\section*{ACKNOWLEDGMENTS}

The authors would like to thank support from National
Natural Science Foundation of China (No. 11135011,
No. 11275125, No. 11221504, No. 11375070, and
No. 11475110), Key Program at Chinese Academy of
Sciences (No. KJCX3-SYW-N2), National Basic Research
Program of China (No. 2015CB856904), Science and
Technology Commission of Shanghai Municipality
(No. 11DZ2260700), Program for Professor of Special
Appointment (Eastern Scholar) at Shanghai Institutions of
Higher Learning, and the ¡°Shu Guang¡± project of Shanghai
Municipal Education Commission and Shanghai Education
Development Foundation.

\appendix
\section{Matching-invariant beta and gamma functions and solutions of the renormalization equations}
\label{app}
A matching-invariant beta function was previously
derived in Ref.~\cite{Peng2006}. The beta and gamma functions
which give matching-invariant running coupling and
running quark masses in QCD can be similarly obtained.
In the present context, the beta and gamma coefficients
are given by
\begin{equation}
\beta_i
=\sum_{k=0}^i\beta_{i,k} N_\mathrm{f}^k, \ \
\gamma_i=\sum_{k=0}^i\gamma_{i,k}N_\mathrm{f}^k
\end{equation}
where the color matrix elements, $\beta_{i,k}$ and $\gamma_{i,k}$, are independent
of the number of flavors $N_\mathrm{f}$, and for
the number of colors $N_\mathrm{c}=3$,
their values can be given, to the 4-loop level, as:
\begin{equation}
[\beta_{i,k}]=
\left[
\begin{array}{cccc}
11/4 & -1/6 &  0  &  0  \\
51/8 & -19/24 & 0 & 0 \\
\frac{2857}{128} & -\frac{4549}{1152} & \frac{79}{1152} & 0 \\
114.230 & -21.5548 & 1.01146 & \frac{23}{1152}
\end{array}
\right]
\end{equation}
and
\begin{equation}
[\gamma_{i,k}]=
\left[
\begin{array}{cccc}
1   & 0  &  0  &  0  \\
101/24 & -5/36 & 0 & 0 \\
1249/64 & -1.30380 & -31/324 & 0 \\
98.9434 & -3.64007 & -0.78090 & \gamma_{3,3}
\end{array}
\right],
\end{equation}
where $\gamma_{3,3}=\zeta_3/36-197/5832\approx -0.00038868$.

In order to have full thermodynamic consistency in the present model,
we need the exact solutions of the renormalization equation (\ref{alphadif})
at a given loop level.
For this purpose, we can apply the approach of variable separation which gives
\begin{equation} \label{RGsep}
\mbox{d}\ln u^2
=\frac{\mathrm{d}\alpha}{\beta(\alpha)}
\equiv \left[
   \acute{\beta}(\alpha)+\frac{\beta_1}{\beta_0\alpha}-\frac{1}{\alpha^2}
       \right]\frac{\mathrm{d}\alpha}{\beta_0},
\end{equation}
 where the acute beta function is defined to be
$
\acute{\beta}(\alpha)
\equiv {\beta_0}/{\beta(\alpha)}-{\beta_1}/({\beta_0\alpha})+{1}/{\alpha^2}.
$
Or, using the expression of the beta function, one has
\begin{equation}
\acute{\beta}(\alpha)
=\frac{\sum_{i=0}^{{\cal N}-2} (\beta_{i+2}-\frac{\beta_1}{\beta_0}\beta_{i+1})\alpha^i-\beta_{\cal N}\alpha^{{\cal N}-2}}
        {\sum_{i=0}^{{\cal N}-1}\beta_i\alpha^i}.
\end{equation}

Integrating both sides of Eq.~(\ref{RGsep}) gives
\begin{equation} \label{alphaeqdaishu}
\ln\frac{u^2}{\Lambda^2}
=\frac{1}{\beta_0\alpha}+\frac{\beta_1}{\beta_0^2}\ln(\beta_0\alpha)
 +\frac{1}{\beta_0^2}W_{\cal N}(\alpha),
\end{equation}
where the function $W_{\cal N}(\alpha)$ is given by
$W_{\cal N}(\alpha)=\beta_0\int_0^{\alpha}\acute{\beta}(x)\,\mbox{d}x$.
For example, it is not difficult to give
\begin{eqnarray}
W_2(\alpha)
 &=&-\beta_1\ln\left(1+\frac{\beta_1}{\beta_0}\alpha\right), \\
W_3(\alpha)
 &=& \frac{2\beta_0\beta_2-\beta_1^2}{\sqrt{4\beta_0\beta_2-\beta_1^2}}
     \arctan\frac{\sqrt{4\beta_0\beta_2-\beta_1^2}}
                 {\beta_1+2\beta_0/\alpha} \nonumber\\
 & & -\frac{\beta_1}{2}\ln\left(\sum_{i=0}^2\frac{\beta_i}{\beta_0}\alpha^i\right).
\end{eqnarray}

The $\Lambda$ in Eq.~(\ref{alphaeqdaishu}) is a QCD scale for the running
coupling. In principle, it is an integration constant. The choice in Eq.~(\ref{alphaeqdaishu})
is to be consistent with the conventional series expansion.
For a chosen $u$, the corresponding $\alpha$ is obtained by numerically solving
the algebraic equation (\ref{alphaeqdaishu}).
At the one-loop level, the function $W_{\cal N}(\alpha)$
becomes $W_1(\alpha)=-\beta_1\ln(\beta_0\alpha)$.
In this case, an explicit solution can be easily obtained, as given
in the first equality of Eq.~(\ref{alphamN1}).

To obtain the strange quark mass, we divide Eq.~(\ref{msdif}) by Eq.~(\ref{alphadif}), giving
\begin{equation} \label{RGms2}
\frac{\mathrm{d}\ln m_s}{\mathrm{d}\alpha}=\frac{\gamma(\alpha)}{\beta(\alpha)}
\equiv\frac{\gamma_0}{\beta_0\alpha}+I_{\cal N}(\alpha),
\end{equation}
where the function $I_{\cal N}(\alpha)$ is defined by
\begin{equation}
I_{\cal N}(\alpha)
\equiv
\frac{\gamma(\alpha)}{\beta(\alpha)}-\frac{\gamma_0}{\beta_0\alpha}
=\frac{\sum_{i=0}^{{\cal N}-2}(\gamma_{i+1}-\frac{\gamma_0}{\beta_0}\beta_{i+1})\alpha^i}
      {\sum_{i=0}^{{\cal N}-1}\beta_i\alpha^i}.
\end{equation}
Integrating Eq.~(\ref{RGms2}) then leads to
\begin{equation} \label{intw}
m_s
=\hat{m}_s\,\alpha^{\gamma_0/\beta_0}
 \exp\left(\int_0^\alpha I_{\cal N}(x)\,\mbox{d}x\right),
\end{equation}
where $\hat{m}_s$ is a QCD mass scale of strange quarks.
 The definite integration in Eq.~(\ref{intw}), i.e.,
 $\int_0^\alpha I_{\cal N}(x)\,\mbox{d}x\equiv w_{\cal N}(\alpha)$,
 can also be analytically carried out, e.g.,
 \begin{eqnarray}
 w_2(\alpha)
 &=& \left(\frac{\gamma_1}{\beta_1}-\frac{\gamma_0}{\beta_0}\right)
     \ln\left(1+\frac{\beta_1}{\beta_0}\alpha\right), \\
 w_3(\alpha)
  &=&
 \frac{\left(
        \frac{\gamma_0}{\beta_0}-2\frac{\gamma_1}{\beta_1}+\frac{\gamma_2}{\beta_2}
       \right)}
     {\sqrt{4\beta_0\beta_2/\beta_1^2-1}}
      \arctan\frac{\sqrt{4\beta_0\beta_2/\beta_1^2-1}}{1+2\beta_0/(\beta_1\alpha)}
 \nonumber\\
  && 
   +\frac{1}{2}\left(\frac{\gamma_2}{\beta_0}-\frac{\gamma_0}{\beta_0}\right)
   \ln\left(\sum_{i=0}^2\frac{\beta_i}{\beta_0}\alpha^i\right).
 \end{eqnarray}

Because $I_1(\alpha)=0$, $w_1(\alpha)$ is also zero.
Accordingly, Eq.~(\ref{intw}) leads to the second equality
in Eq.~(\ref{alphamN1}) at the one-loop level.

\end{document}